\documentclass[12pt]{article}
\usepackage{epsf, amssymb}
\usepackage{epsfig}

\makeatletter
\renewcommand\section{\@startsection {section}{1}{\z@}%
                                   {-3.5ex \@plus -1ex \@minus -.2ex}%nn
                                   {2.3ex \@plus.2ex}%
                                   {\normalfont\large\bfseries}}
\renewcommand\subsection{\@startsection{subsection}{2}{\z@}%
                                     {-3.25ex\@plus -1ex \@minus -.2ex}%
                                     {1.5ex \@plus .2ex}%
                                     {\normalfont\bfseries}}
\makeatother
\def\IZ{\relax\ifmmode\mathchoice
{\hbox{\cmss Z\kern-.4em Z}}{\hbox{\cmss Z\kern-.4em Z}}
{\lower.9pt\hbox{\cmsss Z\kern-.4em Z}} {\lower1.2pt\hbox{\cmsss
Z\kern-.4em Z}}\else{\cmss Z\kern-.4em Z}\fi}
\def\IR{\relax{\rm I\kern-.18em R}}

\def\one{{\hbox{ 1\kern-.8mm l}}}

\newlength{\bredde}
\def\slash#1{\settowidth{\bredde}{$#1$}\ifmmode\,\raisebox{.15ex}{/}
\hspace*{-\bredde} #1\else$\,\raisebox{.15ex}{/}\hspace*{-\bredde}
#1$\fi}

\newsavebox{\zzzbar}
\sbox{\zzzbar}
  {\setlength{\unitlength}{0.9em}
  \begin{picture}(0.6,0.7)
  \thinlines
  \put(0,0){\line(1,0){0.6}}
  \put(0,0.75){\line(1,0){0.575}}
  \multiput(0,0)(0.0125,0.025){30}{\rule{0.3pt}{0.3pt}}
  \multiput(0.2,0)(0.0125,0.025){30}{\rule{0.3pt}{0.3pt}}
  \put(0,0.75){\line(0,-1){0.15}}
  \put(0.015,0.75){\line(0,-1){0.1}}
  \put(0.03,0.75){\line(0,-1){0.075}}
  \put(0.045,0.75){\line(0,-1){0.05}}
  \put(0.05,0.75){\line(0,-1){0.025}}
  \put(0.6,0){\line(0,1){0.15}}
  \put(0.585,0){\line(0,1){0.1}}
  \put(0.57,0){\line(0,1){0.075}}
  \put(0.555,0){\line(0,1){0.05}}
  \put(0.55,0){\line(0,1){0.025}}
  \end{picture}}

\newcommand{\ena}{\end{eqnarray}}
\newcommand{\beqa}{\begin{eqnarray}}
\newcommand{\eeqa}{\end{eqnarray}}
\newcommand{\bea}{\begin{eqnarray}}
\newcommand{\eea}{\end{eqnarray}}

\newcommand{\eq}[1]{(\ref{#1})}

\newcommand{\be}{\begin{equation}}
\newcommand{\ee}{\end{equation}}

\usepackage{graphicx}

\newcommand{\beq}{\begin{equation}}
\newcommand{\eeq}{\end{equation}}
\newcommand{\ber}{\begin{array}}
\newcommand{\eer}{\end{array}}

\newcommand{\del}{\partial}

\newcommand{\dsty}{\displaystyle}

\newcommand{\de}{\delta}

\newcommand{\eps}{\varepsilon}

\usepackage{amsmath}
\usepackage{amssymb}
\newcommand{\tld}{\tilde}
\newcommand{\br}[1]{\left(#1\right)}
\newcommand{\bbr}[1]{\left[#1\right]}

\begin{document}
\begin{titlepage}
\begin{flushright}
arXiv:YYMM.NNNN
\end{flushright}
\vfill
\begin{center}
{\LARGE\bf Quantum evolution across singularities:\vspace{2mm}\\
the case of geometrical resolutions}    \\
\vskip 10mm
{\large Ben Craps,$^a$ Frederik De Roo$^{a,b,}$\footnote{Aspirant FWO} and Oleg Evnin$^a$}
\vskip 7mm
{\em $^a$ Theoretische Natuurkunde, Vrije Universiteit Brussel and\\
The International Solvay Institutes\\ Pleinlaan 2, B-1050 Brussels, Belgium}
\vskip 3mm
{\em $^b$ Universiteit Gent, IR08\\Sint-Pietersnieuwstraat 41, B-9000 Ghent, Belgium}
\vskip 3mm
{\small\noindent  {\tt Ben.Craps@vub.ac.be, fderoo@tena4.vub.ac.be, eoe@tena4.vub.ac.be}}
\end{center}
\vfill

\begin{center}
{\bf ABSTRACT}\vspace{3mm}
\end{center}

We continue the study of time-dependent Hamiltonians with an isolated singularity in their time dependence, describing propagation on singular space-times. In previous work, two of us have proposed a ``minimal subtraction'' prescription for the simplest class of such systems, involving Hamiltonians with only one singular term. On the other hand, Hamiltonians corresponding to geometrical resolutions of space-time tend to involve multiple operator structures (multiple types of dependence on the canonical variables) in an essential way.

We consider some of the general properties of such (near-)singular Hamiltonian systems, and further specialize to the case of a free scalar field on a two-parameter generalization of the null-brane space-time. We find that the singular limit of free scalar field evolution exists for a discrete subset of the possible values of the two parameters. The coordinates we introduce reveal a peculiar reflection property of scalar field propagation on the generalized (as well as the original) null-brane. We further
present a simple family of pp-wave geometries whose singular limit is a light-like hyperplane (discontinuously) reflecting the positions of particles as they pass through it.

\vfill

%\begin{flushleft}
%PACS 11.25.-w, 04.65.+e
%\end{flushleft}
\end{titlepage}
%%%%%%%%%%%%%%%%%%%%%%%%%%%%%%%%%%%%%%%%%%%%%%%%%%%%%%%%%%%%%%%%%%%%%%%%%%%%%%%%%%%%%%%%%%%%%%%%%%%%%%%%%%%%%%%%%%%%%%%%%%%%%%%%%%%%%%%%%%%%%%%%%%%%%%%%%%%%%%%%%%%%%%%%%%%%%%%%%%%%%%%%%%%%%%%%%%%%%

\section{Introduction}

Defining dynamical transitions through space-time singularities entails a very large amount of ambiguity, both technically and conceptually, and, clearly, a deeper and more systematic understanding
of gravitational physics is needed in order to address the issue with full legitimacy. Nevertheless, even in the absence of such understanding, it appears desirable to explore the range of possibilities presented by the problem of evolution across singularities.

In several string theory approaches, holography maps the study of cosmological singularities to the study of quantum mechanics or quantum field theory with certain singular features: couplings may develop a singularity as a function of time (\cite{ben1} and related work), the quantum field theory may live on a singular spacetime (\cite{ben1} and related work) or it may have a potential unbounded below \cite{hertog}. In particular, the model of \cite{ben1} is described by a quantum field theory of matrices on (the future cone of) the compactified Milne space-time. A similar model studied in \cite{robbins} involves quantum field theory of matrices on the nullbrane space-time \cite{FigueroaO'Farrill:2001nx} and its singular limit, the parabolic orbifold \cite{Horowitz:1991ap}. The models presented in \cite{miaoli} are quantum mechanics models of matrices, making them simpler from some points of view. An important question is what happens in these field theories when the space-times they live on develop a singularity. While certain subtle questions related to the large $N$ limit ($N$ being the size of the matrices) have not been fully addressed yet, these models clearly motivate the study of field theory with (near-)singular Hamiltonians.

The parabolic orbifold \cite{Liu:2002ft}, the nullbrane \cite{Liu:2002kb} as well as the compactified Milne space-time have also been studied as (a part of) backgrounds of gravitational theories, in particular as (a part of) string theory space-times. Most attempts to study the singular space-times among those just mentioned in string perturbation theory have failed because both the parabolic orbifold and the Milne orbifold exhibit divergences signaling large gravitational backreaction \cite{Liu:2002ft, Lawrence:2002aj, Horowitz:2002mw, Berkooz:2002je}. See, however, \cite{McGreevy:2005ci} for a closely related model in which the singularity is replaced by a phase with a condensed winding tachyon within perturbative string theory. Note that in the models we shall discuss, these singular space-times host a holographically dual quantum field theory, {\em not} a gravitational theory. In fact, in the present paper we limit ourselves to the study of free field propagation on a fixed (possibly singular) space-time background. Interesting questions related to possible non-gravitational backreaction are postponed to future work.

In a previous publication \cite{qsing}, two of us noted that evolution across space-like or light-like singularities appears to be often described by quantum time-dependent Hamiltonians with an isolated singularity in their time dependence. We then exposed the most conservative way to define a unitary quantum evolution corresponding to such Hamiltonians by modifying the singular time dependences to become distributions while keeping the operator structure of the Hamiltonian unchanged. This approach is relevant when the transition through the singularity is dominated by a single term (single operator structure) in the Hamiltonian, and one can think of this way to define the transition through the singularity as a sort of ``minimal subtraction''. In the absence of further physical specifications, this procedure appears to be the most natural way to define evolution across singularities.

However, in many geometrical contexts, another approach appears to be more natural. Namely, one may want to resolve the singular geometry into a smooth space, and then try to take
the singular limit in such a way that the dynamical evolution remains well-defined. This is non-trivial, since na\"\i ve resolutions of a singular space-time will generically not lead to a well-defined dynamics in the singular limit. As we shall argue shortly, constructing such geometrical resolutions will typically take us outside the formal scope of \cite{qsing}, and it is the aim of our present paper to investigate what kind of mathematical structures the geometrical resolutions of space-like (light-like) singularities involve: in general, as well as in a few specific examples, which will be at the focus of our attention.

In \cite{qsing}, we divised a kind of ``minimal subtraction'' prescription and described its particular implementation for the case of a free scalar field propagating on the compactified Milne universe. Here, we would like to argue that the approach of \cite{qsing} does not lend itself to a geometrical interpretation, and, therefore, should one be interested in geometrical resolutions of space-time singularities, a more general framework is required.

To recapitulate briefly, the metric of the compactified Milne universe is
\be\label{MilneMetric}
ds^2=-dt^2+t^2dx^2,\ \ \ \ \ \ x\sim x+2\pi,
\ee
and the corresponding free scalar field Hamiltonian is
\be
H=\frac1{2|t|}\int dx\,\left(\pi_\phi^2+{\phi^\prime}^2\right)+\frac{m^2|t|}2\int dx\, \phi^2.
\label{hmsingular}
\ee
With this form of the Hamiltonian, the Schr\"odinger equation cannot be integrated through $t=0$ on account of the singularity of $1/|t|$.

The idea of the ``minimal subtraction'' scheme of \cite{qsing} is to keep the operator structure of the Hamiltonian unchanged and to modify the singular time dependences in (\ref{hmsingular}) locally at $t=0$ by subtracting terms proportional to (possibly) resolved $\delta$-functions and its derivatives such that the time dependences become well-defined in the sense of distributions\footnote{This procedure bears a strong formal resemblance to the conventional renormalization of local field theories by subtracting local counter-terms, and it can be thought of (see \cite{qsing} for further discussion) as renormalizing the time dependence of (\ref{hmsingular}).}. Then, the Schr\"odinger equation can be integrated. Put differently, one can replace $1/|t|$ in (\ref{hmsingular}) by its regulated version $f_{1/|t|}(t,\eps)$ (with $\eps$ being a regularization parameter), in such a way that, as $\eps$ is taken to 0, $f_{1/|t|}(t,\eps)$ converges to a distribution ${\cal F}_{1/|t|}(t)$, and this distribution ${\cal F}_{1/|t|}(t)$ equals $1/|t|$ everywhere away from $t=0$. A possible choice is
\be
f_{1/|t|}(t,\eps)={1\over\sqrt{t^2+\epsilon^2}}+2\ln(\mu\epsilon){\epsilon\over\pi(t^2+\epsilon^2)},\label{reginvabs}
\ee
with $\mu$ an arbitrary mass scale.

With this approach, one obtaines a regularized version of the Hamiltonian (\ref{hmsingular}), namely
\be
H=\frac1{2}\,f_{1/|t|}(t,\eps)\int dx\,\left(\pi_\phi^2+{\phi^\prime}^2\right)+\cdots
\label{hmreg}
\ee
such that, as $\eps$ is taken to 0, the evolution away from $t=0$ becomes identical to that arising from (\ref{hmsingular}), and, furthermore, the system displays a well-defined (unitary) transition through $t=0$.

The ``minimal subtraction'' procedure we have just briefly re-stated, is a consistent evolution prescription in itself. However, a direct inspection of (\ref{hmreg}) shows that the regularized version of our dynamics does not admit a geometrical interpretation (nor should one think of its singular limit, albeit well-defined, as being geometrical).

\begin{figure}
\centering
\epsfig{file=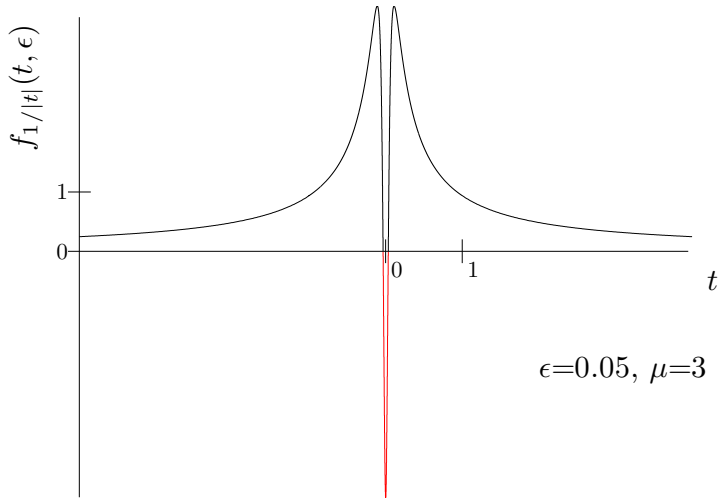, width=4in}
\caption{Negative contribution around $t=0$ in regularised $f_{1/|t|}(t,\epsilon)$ (\ref{reginvabs})}
\label{fig:RegInvAbsRed}
\end{figure}

The problem with constructing a geometrical interpretation of (\ref{hmreg}) is that, since $f_{1/|t|}(t,\eps)$ has an $\eps\to 0$ limit as a distribution, the $\eps\to 0$ limit of
\be
\int\limits_{-t_0}^{t_0}dt\,f_{1/|t|}(t,\eps)
\label{int}
\ee
must exist. Furthermore, as stated above, the $\eps\to 0$ limit of $f_{1/|t|}(t,\eps)$ must equal $1/|t|$ everywhere away from $t=0$. For that reason, in order for the limit of (\ref{int}) to exist, $f_{1/|t|}(t,\eps)$ should be very large and {\it negative} somewhere in the $\eps$-neighborhood of $t=0$ so that the positive divergence from integrating $1/|t|$ is compensated in (\ref{int}). This is clearly apparent in figure \ref{fig:RegInvAbsRed}. However, the coefficient of the kinetic term in the Hamiltonian of a field in a geometrical background comes from the square root of the determinant of the metric (and the coefficients of the inverse metric), and it needs to be {\it positive} (as is the function $1/|t|$ appearing in (\ref{hmsingular})).

For that reason, there appears to be a conflict between the demands of positivity for certain coefficients in the Hamiltonian arising if one pursues a geometrical interpretation, and negative contributions introduced by our ``minimal subtraction'' recipe. If one is to construct a geometrical resolution of dynamics on a singular space-time background, one generally needs to relax the specifications of the ``minimal subtraction'' approach, and permit modifications in the operator structure of the Hamiltonian, as well as its time dependence, in the vicinity of the singular region. One will then typically end up with a situation where a few different operator structures in the Hamiltonian essentially contribute to the transition to the singular region:
\be
H(t)=\sum\limits_i f_i(t,\eps)  H_i,
\label{multiop}
\ee
where $H_i$ are time-independent operators and $f_i$ are time-dependent number-valued functions. $\eps$ is a regularization parameter, and the implication is that, as $\eps$ is taken to 0, some of the $f_i$'s may develop isolated singularities at a certain value of $t$, which we shall choose to be $t=0$. It is the commutation properties of those different terms in the Hamiltonian that are responsible for divergence cancellation (rather than explicit negative contributions introduced through the ``minimal subtraction'' scheme of \cite{qsing}).

In what follows, we shall review some relevant properties of quantum time-dependent Hamiltonians involving multiple operator structures, and proceed with applying this range of techniques to constructing singular limits of dynamics in a few different geometrical backgrounds related to the null-brane space-time. In section~2, we start by discussing the
evolution properties for Hamiltonians of the form \eq{multiop}. In section~3, we examine as an example the null-brane, which is the geometric
resolution (through a parameter $R$, which plays the role of $\epsilon$ in this case) of the parabolic orbifold. We introduce
a generalized null-brane metric depending on continuous parameters $(R,\alpha,\beta)$. The propagation
of a free scalar field on this space-time will be described by computing its mode functions, derived by solving
the wave equation exactly using WKB methods. Then we discuss properties of the mode functions
of a free scalar field in the singular limit $R \to 0$. We shall find that the limit of the mode functions exists for certain discrete values of the parameters $\alpha$ and $\beta$. We give a qualitative discussion of the results in subsection~3.3. In section~4, we introduce the light-like reflector plane, a simple yet curious space-time that can be used as a toy model for light-like space-time singularities. Section~5 contains our conclusions. Appendix~A contains a discussion of the minimal subtraction prescription of \cite{qsing} applied to the parabolic orbifold. In appendix~B, we give some technical details related to focusing properties and Maslov phases.

%%%%%%%%%%%%%%%%%%%%%%%%%%%%%%%%%%%%%%%%%%%%%%%%%%%%%%%%%%%%%%%%%%%%%%%%%%%%%%%%%%%%%%%%%%%%%%%%%%%%%%%%%%%%%%%%%%%%%%%%%%%%%%%%%%%%%%%%%%%%%%%%%%%%%%%%%%%%%%%%%%%%%%%%%%%%%%%%%%%%%%%%%%%%%%%%%%%%%

\section{Time-dependent Hamiltonians involving multiple operator structures}

In preparation for our analysis of geometrical resolutions and their singular limits, we shall review the quantum dynamics described by Hamiltonians of the form \eq{multiop}. Our ultimate question will be whether the $\eps\to 0$ limit of the evolution operator corresponding to (\ref{multiop}) exists.

It is in general impossible to solve the Schr\"odinger equation corresponding to the Hamiltonian (\ref{multiop}). The familiar symbolic solution for the evolution operator $U(t_1,t_2)$ involves the time-ordering symbol ${\mbox T}$:
\be
U(t_1,t_2)={\mbox T}\left[-i\int\limits_{t_1}^{t_2} dt\, H(t)\right].
\label{timeorder}
\ee

The above representation can be further transformed in an instructive way using a technique known as the Magnus expansion \cite{magnus}. The operator $U$ belongs to the group of unitary operators on the Hilbert space, and the Magnus expansion can be thought of as an analog of the Baker-Campbell-Hausdorff formula for finite-dimensional Lie groups (the latter is discussed in many textbooks on group theory, for example, in \cite{hall}). The expansion can be symbolically written as:
\be
\begin{array}{l}
\dsty U(t_1,t_2)=\exp\left[-i\int\limits_{t_1}^{t_2}dt\,H(t)+\eta_1\int dt\,dt'\,[H(t),H(t')]\right.\vspace{2mm}\\
\dsty\hspace{4cm}\left.+i\,\eta_2
\int dt\,dt'\,dt''\,[H(t),[H(t'),H(t'')]]+\cdots\right],
\end{array}
\label{magnus}
\ee
with some numerical coefficients $\eta_1$, $\eta_2,\ldots$ (their values will not be important for us, and it appears they can only be derived recursively \cite{magnus}). The key property of the above expression is that the higher order terms are entirely expressed through higher order nested commutators of $H(t)$ at different moments of time.

Even though, in a completely general setting, the Magnus expansion is hopelessly intractable, it displays the broad range of opportunities for divergence cancellation in a singular limit of the dynamics described by (\ref{multiop}). Namely, for the case of (\ref{multiop}), the Magnus expansion (\ref{magnus}) will contain all kinds of combinations of the $f_i$ and their products, in such a way that, even if $f_i$ develop very strong singularities as $\eps$ is taken to 0, the limit of $U(t_1,t_2)$ may still exist. For example, even if all $f_i$ are positive, cancellations may still occur on account of the commutation properties of $H_i$.

Should such cancellations take place, one may think of the $\eps\to 0$ limit of (\ref{multiop}) as an operator-valued generalization of conventional distributions: just as ordinary distributions may contain singularities in a way that permits evaluating ordinary integrals, the Hamiltonian (\ref{multiop}) will contain singularities in a way that permits evaluating the time-ordered exponential integral in (\ref{timeorder}). A systematic exploration of such generalized operator-valued ``distributions'' may be interesting to pursue, but lies outside of the scope of the present paper.

There is a special case when the above analysis can be taken significantly further. Namely, it may turn out that, for all moments of time, the operator $U$ of (\ref{timeorder})  belongs to a finite-dimensional subgroup of the unitary group of the Hilbert space. This situation has been described as a presence of a {\it dynamical group} (see \cite{zhengfenggilmore,malkinmanko} and references therein). For the Hamiltonians of the form (\ref{multiop}), there will exist a finite-dimensional dynamical group if the set of nested commutators of $H_i$'s closes on a finite-dimensional linear space of operators (which would serve as the Lie algebra of the dynamical group). Should that happen, one would be able to use the the closed resummed version of the Baker-Campbell-Hausdorff formula for finite-dimensional Lie groups (see, for example, \cite{hall}) to treat the Magnus expansion, or, alternatively, the Schr\"odinger equation can be reduced to a finite number of ordinary differential equations\footnote{The analytic power of the dynamical group approach does not appear completely clear or fully explored. It certainly does apply to all linear quantum systems; however, in that case, the conventional WKB analysis would suffice. Beyond linear systems, the relevant finite-dimensional subalgebras of Hermitean operators may be difficult to construct and/or classify. Nevertheless, some non-trivial examples of dynamical groups for quantum-mechanical systems do exist (see, for example, \cite{dodonovmankorosa}).} describing the evolution on the finite-dimensional dynamical group manifold \cite{zhengfenggilmore,malkinmanko,yuen}. In practical terms, one can choose a particular low-dimensional faithful linear representation of the dynamical group furnished by matrices $M$, and write down the Schr\"odinger equation in this representation:
\be
i\frac{dM(t,t_0)}{dt}=\varphi(H(t))M,\qquad M(t_0,t_0)=1,
\ee
where $\varphi$ is a homomorphism from Hilbert space operators onto the representation furnished by $M$.
(This is a finite-dimensional system of ordinary differential equations.) Given the solution for $M(t,t_0)$, one can reconstruct the original evolution operator as $\varphi^{-1}(M(t,t_0))$.

While the specific examples of quantum dynamics discussed in this paper will be constructed using a somewhat unconventional application of WKB methods, the fact this ``double-semiclassical'' analysis is possible reflects an underlying dynamical group structure inherent to the systems we shall work with. After explicitly computing the mode functions encoding the dynamics, we shall investigate (in this greatly simplified setting) the existence of singular limits. It will then be possible to circumnavigate the formal complications introduced by the non-commuting structures in (\ref{multiop}), and examine the limiting case of evolution on a singular space-time background.

%%%%%%%%%%%%%%%%%%%%%%%%%%%%%%%%%%%%%%%%%%%%%%%%%%%%%%%%%%%%%%%%%%%%%%%%%%%%%%%%%%%%%%%%%%%%%%%%%%%%%%%%%%%%%%%%%%%%%%%%%%%%%%%%%%%%%%%%%%%%%%%%%%%%%%%%%%%%%%%%%%%%%%%%%%%%%%%%%%%%%%%%%%%%%%%%%%%%%%

\section{Generalized null-brane space-times and singular limits}

In this section, we study quantum dynamics of a free scalar field propagating on the parabolic orbifold and some geometrical resolutions thereof, including the null-brane.

\subsection{The null-brane geometry, the parabolic orbifold and generalizations}

The null-brane spacetime, originally introduced in \cite{FigueroaO'Farrill:2001nx}, was studied in the context of perturbative string theory in \cite{Liu:2002kb, Fabinger:2002kr, RS}; a matrix theory description was provided in \cite{robbins}.

Consider Minkowski space-time in light-cone coordinates $ds^2=-2dx^+dx^-+dx^2+dz^2$. The null-brane is obtained by identifying
\begin{eqnarray}
\begin{bmatrix}x^+\\x\\x^-\\\end{bmatrix} \sim \mathrm{exp}(2\pi n \mathcal{J}) \begin{bmatrix}x^+\\x\\x^-\\\end{bmatrix}\,,
 \hspace{10mm} z \sim z+2\pi n R\,,\hspace{10mm} n \in \mathbb{Z}\,,\hspace{10mm}\mathcal{J}=\begin{bmatrix}0&0&0\\1&0&0 \\ 0&1&0\\\end{bmatrix}.
\label{nb_identification}
\end{eqnarray}

In the $R\rightarrow0$ limit the null-brane reduces to the parabolic orbifold \cite{Horowitz:1991ap, Liu:2002ft} times the line labeled by $z$. In this sense, the null-brane is a geometrical regularization of the parabolic orbifold.
In Rosen coordinates, the metric on the parabolic orbifold is given by
\begin{equation}
ds^2=-2dy^+dy^-+(y^+)^2 dy^2.\label{metricRosenPO}
\end{equation}
In \cite{Liu:2002kb}, the null-brane geometry was discussed in two coordinate systems,
\bea
ds^2&=&-2dy^+dy^-+du^2+(R^2+(y^+)^2) dy^2+2R dy du;\label{metricRosen}\\
ds^2&=&-2d\tld{x}^+d\tld{x}^-+d\tld{x}^2+(R^2+\tld{x}^2) d\theta^2+2(\tld{x}^+d\tld{x}-\tld{x}d\tld{x}^+)d\theta,\label{metricTilde}
\eea
related to Minkowski coordinates by
\bea\label{relCoord}
&&x^+=y^+,\;x=y y^+,\;x^-=y^-+\frac{1}{2}y^+ y^2,\;u=z-Ry;\\
&&x^+=\tld{x}^+,\;x=\tld{x}+\theta \tld{x}^+,\;x^-=\tld{x}^-+\theta\tld{x}+\frac{1}{2}\theta^2\tld{x}^+,\;\theta=\frac{z}{R}.
\eea
Unfortunately, neither coordinate system is fully satisfactory for studying the $R\to 0$ limit of dynamics on the null-brane.%
\footnote{
Following the discussion in the previous section, with $R$ playing the role of $\epsilon$, we would like to phrase the dynamics in terms of a Hamiltonian that has the structure $H=\sum_i f_i(t,R)H_i$ in which $f_i(t,R)$ is regular in $t$ for $R\neq 0$ and regular away from $t=0$ for $R=0$. The terms that appear in the Hamiltonian can be easily deduced from the inverse metric.
}
On the one hand, the $y\text{-coordinates}$ are not globally defined since they are singular at $y^+=0$ for any $R$. On the other hand, the $\tld{x}\text{-coordinates}$, which are nonsingular for $R\neq0$, do not have an $R\to 0$ limit even away from the parabolic orbifold singularity, as the determinant of the metric is $-R^2$ everywhere.

Therefore, we now introduce new coordinates that interpolate between the $\tld{x}\text{-coordinates}$ (for small $\tld{x}^+$) and the $y\text{-coordinates}$ (for large $y^+$):
\bea
\label{LMScotrafo}
X^+ &=&\tld{x}^+ =y^+;\\
X^-&=&\tld{x}^--\frac{1}{2}\frac{\tld{x}}{\tld{x}^+}\br{1-\frac{R^4}{\br{R^2+(\tld{x}^+)^2}^2}}=y^-+\frac{R^2}{2}\frac{y^+ u^2}{\br{R^2+(y^+)^2}^2};\nonumber\\
X&=&-\frac{R \tld{x}}{\sqrt{R^2+(\tld{x}^+)^2}}=\frac{y^+ u}{\sqrt{R^2+(y^+)^2}};\nonumber\\
\Theta&=&\theta+\frac{\tld{x}}{\tld{x}^+}\br{1-\frac{R^2}{R^2+(\tld{x}^+)^2}}=y+\frac{R u}{R^2+(y^+)^2}.\nonumber
\eea
In this coordinate system, the metric has determinant $-(R^2+(X^+)^2)$.

The metric of the null-brane space-time, written in our new coordinates, can be naturally generalized to a two-parameter family of metrics, which we will call \emph{generalized null-brane}. The family is labeled by parameters $\alpha$ and $\beta$, the original null-brane corresponding to $\alpha=3, \beta=2$:
\begin{equation}
ds^2=\frac{R^2 X^2 \br{\beta^2-\alpha}}{\br{R^2+(X^+)^2}^2}\br{dX^+}^2\,-2 dX^+ dX^-+ \frac{2\beta R X}{\sqrt{R^2+(X^+)^2}}dX^+ d\Theta+\br{R^2+(X^+)^2} d\Theta^2+dX^2.\label{lineElementGNB}
\end{equation}

%%%%%%%%%%%%%%%%%%%%%%%%%%%%%%%%%%%%%%%%%%%%%%%%%%%%%%%%%%%%%%

\subsection{Free scalar field on the generalized null-brane}

\subsubsection{General dynamical preliminaries}

We consider a free scalar field on the generalized null-brane metric. After Fourier transforming with respect to $X^-$ and $\Theta$,
\begin{equation}
\phi(X^+,X^-,X,\Theta)=\frac{1}{2 \pi} \sum_{k_{\Theta}} \int dk_- \phi_{k_-,k_{\Theta}} \mathrm{exp}\br{i k_- X^- + i k_{\Theta} \Theta},\label{fourierTransf}
\end{equation}
and suppressing the indices $k_-$ and $k_{\Theta}$, the action reads
\bea
S&=&\sum_{k_{\Theta}}  \int \mathrm{d}X^+\,\mathrm{d}k_-\,\mathrm{d}X\, \,\,\sqrt{R^2+(X^+)^2}\Biggl[\frac{i k_-}{2} \bigl(\phi \partial_{X^+} \phi^*-\phi^*\partial_{X^+} \phi\bigr)-\frac{\partial_{X}\phi\partial_{X}\phi^*}{2}\nonumber\\
&&\ \ \ \ \ \ \ \ \ -\biggl(\frac{m^2}{2}+\frac{k_{\Theta}^2}{ 2 (R^2+(X^+)^2)}+\frac{\alpha X^2 R^2 k_-^2}{2 (R^2+(X^+)^2)^2}+\frac{k_{\Theta} k_-\beta X R}{(R^2+(X^+)^2)^{3/2}}\biggr)\phi\phi^*\Biggr].\label{actionMSF}
\eea
Denoting $\partial_{X^+}\phi$ as $\dot{\phi}$, the wave equation reads
\bea\label{waeq}
-i\dot{\phi}&=&\frac{i X^+}{2 \br{R^2+(X^+)^2}}\phi-\frac{\partial^2_X\phi}{2 k_-}+\frac{\beta X R k_{\Theta}}{\bigl(R^2+(X^+)^2\bigr)^{3/2}} \phi\\
&&\hspace{4mm}+\frac{k_{\Theta}^2}{2 k_- \br{R^2+(X^+)^2}}\phi+\frac{\alpha}{2}\frac{X^2 R^2 k_-}{\bigl(R^2+(X^+)^2\bigr)^2}\phi+ \frac{m^2}{2 k_-}\phi.\nonumber
\eea
As the Lagrangian is first order in time derivatives, we have to deal, in principle, with constraints when deriving the corresponding Hamiltonian. One can take a shortcut interpreting $\phi$ as a canonical coordinate and $\pi\equiv i k_- \sqrt{R^2+(X^+)^2}\phi^*$ as its conjugate momentum. The Hamiltonian reads
\bea
H &=&\sum_{k_{\Theta}}\sum_{k_-\not=0}\int\,\mathrm{d}X\,\pi \Biggl[\frac{X^+}{2 \br{R^2+(X^+)^2}}-\frac{i m^2}{2 k_-} + \frac{i}{2k_-} \partial^2_X-\frac{i \beta X R k_{\Theta}}{\bigl(R^2+(X^+)^2\bigr)^{3/2}}\nonumber\\
&&\ \ \ \ \ \ \ \ \ \ \ \ \ \ \ \ \ -\frac{i}{2 k_-}\frac{k_{\Theta}^2}{R^2+(X^+)^2}
-\frac{i \alpha}{2}\frac{X^2 R^2 k_-}{\bigl(R^2+(X^+)^2\bigr)^2}\Biggr]\phi.\label{ham_piphi}
\eea
It is manifestly of the form $H=\sum_i f_i(t,R)H_i$, in other words,
it belongs to the class of Hamiltonians we single out in section~2.

We now show that the Hamiltonian (\ref{ham_piphi}) leads to a finite-dimensional {\it dynamical group} structure of the type discussed in section~2. The canonical variables $\pi(X,t)$ and $\phi(X,t)$ appear in four combinations $\int \pi\,\partial_X^2 \phi,\;\int\pi\phi,\;\int\pi X \phi\;\text{and}\;\int\pi X^2 \phi$. The commutation relations for such operators are given by
\begin{equation}
\left[\int\pi\,\hat{A}(X)\,\phi\,dX\,,\;\int\pi\,\hat{B}(X)\,\phi\,dX\right]=\int\pi[\hat{A},\hat{B}]\phi\,dX,\label{commrel}
\end{equation}
reducing the commutator algebra to that of $\{\partial^2_X,\,1,\,X,\,X^2\}$, which closes after the addition of $\{X\partial_X,\,\partial_X\}$. Equivalently, one can form the standard (single degree of freedom) creation and annihilation operators $a$ and $a^\dagger$ out of $X$ and $\del_X$ to conclude that the Lie algebra of the dynamical group
is spanned by $n=a^{\dagger}a+1/2,a^{\dagger 2},a^2,a^{\dagger},a$ and $I$. Given the inclusion of powers of the creation operator up to $a^{\dagger 2}$, it is not surprising that the corresponding algebra
has become known as the {\it two-photon algebra}, or $h_6$, and has
been featured in discussions of quantum optics, and squeezed states
in particular (see, for example \cite{zhengfenggilmore}). A complete
formal analysis of quantum dynamics on the two-photon group has been given in \cite{yuen}. 

Following the general picture presented in section 2, we could use the two-photon group considerations of \cite{yuen} to reduce the question of
free scalar field dynamics on the generalized null-brane to ordinary
differential equations. In our present setting, however, one
can perform these operations in a considerably more familiar guise.
Namely, since the free scalar field is linear, solving for its
quantum dynamics amounts to constructing a complete set of solutions
to the classical wave equation. Furthermore, the classical wave equation
turns out to be equivalent to the Schr\"odinger equation for a {\it linear} auxiliary one-dimensional quantum system. Because the auxiliary system is linear, its Schr\"odinger equation (i.e., the wave equation
of the original scalar field) can be solved exactly by WKB methods.
The latter effectively reduce the problem to ordinary differential equations (the classical equations of motion of the auxiliary linear system). Thus, one attains the same level of simplification as
one would through performing the analysis of \cite{yuen}. One can
refer to the above procedure as ``double-semiclassical'' analysis (there is an (exact) WKB procedure leading from a free quantum scalar field to the wave equation for the mode functions, and an (exact) WKB procedure leading from the wave equation for the mode functions to a one-dimensional auxiliary classical system). Note that both the ``double-semiclassical'' approach and the general dynamical group approach of \cite{yuen} (which are essentially one and the same thing)
are made possible by the fact that the metric of the generalized null-brane is a quadratic polynomial in the $X$-variable.

For linear quantum systems, it is most common to work in the Heisenberg picture, instead of (equivalently) deriving WKB wave functions in the Schr\"odinger picture. One obtains the solution for the Heisenberg field operator as an expansion in terms of a complete set of mode functions $u_y(X,X^+)$ (with $y$ being a generic basis label) satisfying the classical equations of motion:
\be
\phi_{k_-,k_{\Theta}}=\int \mathrm{d}y \;u_y(X,X^+)a(y).
\ee
The corresponding conjugate momentum is
\be
\pi_{k_-,k_{\Theta}}=i k_- \sqrt{R^2+\br{X^+}^2} \int \mathrm{d}y \;u^*_y(X,X^+)a^{\dagger}(y).
\ee
If one demands the standard commutation relations for the creation-annihilation operators $a^\dagger$ and $a$, the canonical commutation relation between $\pi$ and $\phi$ determine the normalisation of the mode functions (this is the analog for first order systems of the Klein-Gordon norm):
\begin{equation}
\delta(X-\tld{X})=k_-\sqrt{R^2+\br{X^+}^2}\int \mathrm{d}y\;u_y^*(X,X^+)u_y(\tld{X},X^+).\label{CCnorm}
\end{equation}

%%%%%%%%%%%%%%
\subsubsection{``Double-semiclassical'' solution of the wave equation}
\label{2-1/2cl}

Denoting $X^+$ by $t$, (\ref{waeq}) takes the form of an auxiliary Schr\"odinger equation with Hamiltonian
\begin{equation}
\mathcal{H}=\frac{i t}{2 \br{R^2+t^2}}+\frac{P^2}{2 k_-}+\frac{\beta X R k_{\Theta}}{\bigl(R^2+t^2\bigr)^{3/2}}+\frac{k_{\Theta}^2}{2 k_- \br{R^2+t^2}}
+\frac{\alpha}{2}\frac{X^2 R^2 k_-}{\bigl(R^2+t^2\bigr)^2}+ \frac{m^2}{2 k_-}\label{hamilt}
\end{equation}
(up to a sign difference in the left hand side of \eq{waeq}).
As the corresponding Hamiltonian (\ref{hamilt}) is quadratic in $X$, \eq{waeq} can be solved exactly by WKB methods. The starting point is the observation that the ansatz
\begin{equation}
\phi(X_1,t_1|X_2,t_2)={\cal A}(t_1,t_2) \mathrm{exp}\br{-i S_{cl}\bbr{X_1,t_1|X_2,t_2}}\label{WKBansatz}
\end{equation}
solves \eq{waeq} (with $t_2\to t$) if
\bea
S_{cl}=\int_{t_1}^{t_2}\,\mathrm{d}t\,\br{P \dot{X} - \mathcal{H}}{\Big|_{X=X_{cl}(X_1,t_1|X_2,t_2)}};\label{WKBactie}\\
-2 k_-\,\frac{\partial {\cal A}(t_1,t)}{\partial t}={\cal A}(t_1,t)\,\frac{\partial^2 S_{cl}\bbr{X_1,t_1|X,t}}{\partial X^2}.\label{WKB_pref}
\eea
Here, $S_{cl}$ is the classical action with boundary conditions $X(t_1)=X_1,\ X(t_2)=X_2$. More general solutions to \eq{waeq} are obtained by integrating \eq{WKBansatz} over $X_1$, weighted by an arbitrary smooth wavepacket.

A subtlety arises in this ansatz when the dynamical evolution reaches a {\it focal point} $t_2=t^*$, where the classical action diverges, unless a certain relation between $X_1$ (``the source'') and $X_2$ (``the image'') is met. At such focal points, the differential equation for ${\cal A}(t_1,t_2)$ becomes singular. In that case, one solves the WKB equations away from $t^*$ and connects the solution by a phase jump at the focal point. The phase jump should be chosen such that convolutions of \eq{WKBansatz} with a smooth wavepacket are continuous across the focal point. The general 
guidelines for this procedure are best familiar in the context
of {\it caustic} submanifolds in geometrical optics (see, for example, \cite{rohmer,maslov}) and the above-mentioned correction pre-factors have become known as the {\it Maslov phases}.
We shall give some further details in appendix~\ref{app_maslov}.

In order to compute the classical action, we first consider the classical equation of motion:
\begin{subequations}\label{classX}
\begin{align}
&\ddot{X}+\alpha\frac{R^2}{\bigl(R^2+t^2\bigr)^2}X=-\frac{\beta k_{\Theta}}{k_-}\frac{R}{\bigl(R^2+t^2\bigr)^{3/2}},\label{eq_X}\\
&X(t_1)=X_1,\qquad X(t_2)=X_2.
\end{align}
\end{subequations}
This equation is actually exactly solvable, and it has become known as
the equation for ``bending of a double-walled compressed bar with a parabolic cross-section'' \cite{zaitsev}. It can be reduced to a driven harmonic oscillator with constant frequency via substitution $X=\sqrt{R^2+t^2} \chi(\eta(t))$, taking $\eta=\mathrm{arctan}(t/R)$:
\begin{equation}
\frac{d^2\chi}{d\eta^2} + \br{1+\alpha}\,\chi=-\frac{\beta k_{\Theta}}{R k_-}.\label{eqchi}
\end{equation}
In order to give a transparent derivation of the solution to (\ref{classX}) and the corresponding value of the classical action, we first consider the two independent solutions to the homogeneous version of (\ref{eq_X}):
\begin{subequations}\label{homSol}
\begin{align}
f(t)=\sqrt{R^2+t^2}\;\mathrm{sin}\br{\sqrt{1+\alpha}\,\mathrm{arctan}\frac{t}{R}},\\
h(t)=\sqrt{R^2+t^2}\;\mathrm{cos}\br{\sqrt{1+\alpha}\,\mathrm{arctan}\frac{t}{R}}.
\end{align}
\end{subequations}
A useful object to consider is the Dirichlet Green function of the operator $\del^2_t+\alpha R^2/\bigl(R^2+t^2\bigr)^2$:
\be
G(t,t'|t_1,t_2)=\frac{\left(f_1h(t_<)-h_1f(t_<)\right)\left(f_2h(t_>)-h_2f(t_>)\right)}{W[f,h](f_1h_2-h_1f_2)},
\label{greenf}
\ee
satisfying
\be
\left(\del^2_{t}+\frac{\alpha R^2}{\bigl(R^2+t^2\bigr)^2}\right)G(t,t'|t_1,t_2)=\de(t-t'),\qquad G(t_1,t'|t_1,t_2)=0,\qquad G(t_2,t'|t_1,t_2)=0
\ee
with $W[f,h]=f\dot{h}-h\dot{f}$ being the Wronskian of $f(t)$ and $h(t)$ (independent of $t$), $t_2>t_1$, $t_<=\mathrm{min}(t,t')$, $t_>=\mathrm{max}(t,t')$ and $f_1=f(t_1)$, $h_1=h(t_1)$, etc.

With the Green function given by (\ref{greenf}), and $b(t)$ denoting the right hand side of equation (\ref{eq_X}), one can write down
the solution to (\ref{classX}) as
\be
\begin{array}{l}
\dsty X_{cl}(t|X_1,t_1;X_2,t_2)=-X_1\del_{t'}G(t,t'|t_1,t_2)\Big|_{t'=t_1}+X_2\del_{t'}G(t,t'|t_1,t_2)\Big|_{t'=t_2}\vspace{2mm}\\
\dsty\hspace{7cm}+\int\limits_{t_1}^{t_2} dt'\,G(t,t'|t_1,t_2)b(t')
\end{array}
\ee

Given the above formulas, the classical action can be written in
a relatively general form that will turn out to be useful later:
\begin{subequations}\label{Scl_general}
\begin{align}
\displaystyle
&S_{cl}=-\frac{k_-}{2}\bbr{\frac{h_2\dot{f}_1-f_2\dot{h}_1}{f_1h_2-h_1f_2}} X^2_1+ \frac{k_-}{2}\bbr{\frac{f_1\dot{h}_2-h_1\dot{f}_2}{f_1h_2-h_1f_2}} X^2_2 -k_-\bbr{\frac{W[f,h]}{f_1h_2-h_1f_2}} X_1X_2\\
&-k_-\int_{t_1}^{t_2}\mathrm{d}t\,\,b(t)\,\br{\frac{h_2f(t)-f_2h(t)}{f_1h_2-f_2h_1}}X_1- k_-\int_{t_1}^{t_2} \mathrm{d}t\,\,b(t)\,\br{\frac{f_1h(t)-h_1f(t)}{f_1h_2-f_2h_1}} X_2\\
&+k_-\int_{t_1}^{t_2}\mathrm{d}t'\int_{t_1}^{t'}\mathrm{d}t\,\,b(t)\frac{\bigl(f_1h(t)-h_1f(t)\bigr)\bigl(f_2h(t')-h_2f(t')\bigr)}{W[f,h] (f_1h_2-h_1f_2)}b(t')\\
&- \frac{m^2}{2 k_-}(t_2-t_1)-\frac{i}{2}\ln\frac{\sqrt{R^2+{t_2}^2}}{\sqrt{R^2+{t_1}^2}}
\end{align}
\end{subequations}
Of course, one can also solve the equations of motion (\ref{classX}) explicitly:
\begin{subequations}
\label{X_solution}
\begin{align}
\displaystyle
X=X_1\frac{\sqrt{R^2+t^2}}{\sqrt{R^2+t_1^2}} \frac{\mathrm{sin}2 \Delta_{t2}}{\,\mathrm{sin}2\Delta_{12}\,}&+X_2\frac{\sqrt{R^2+t^2}}{\sqrt{R^2+t_2^2}}\frac{\mathrm{sin}2 \Delta_{1t}}{\,\mathrm{sin}2\Delta_{12}\,}\nonumber\\&
-\frac{\beta k_{\Theta}\sqrt{R^2+t^2}}{Rk_-(1+\alpha)}\biggl[1-\frac{\mathrm{sin}2 \Delta_{t2}}{\,\mathrm{sin}2\Delta_{12}\,}-\frac{\mathrm{sin}2 \Delta_{1t}}{\,\mathrm{sin}2\Delta_{12}\,}\biggr]\label{solX}\\
\Delta_{12}=\frac{\sqrt{1+\alpha}}{2}&\br{\mathrm{arctan}\frac{t_2}{R}-\mathrm{arctan}\frac{t_1}{R}}\\
\Delta_{t2}=\frac{\sqrt{1+\alpha}}{2}&\br{\mathrm{arctan}\frac{t_2}{R}-\mathrm{arctan}\frac{t}{R}}\\
\Delta_{1t}=\frac{\sqrt{1+\alpha}}{2}&\br{\mathrm{arctan}\frac{t}{R}-\mathrm{arctan}\frac{t_1}{R}}
\end{align}
\end{subequations}
The classical action can now be evaluated either by brute force using the explicit classical solution (\ref{X_solution}), or, with less work, from (\ref{Scl_general}):
\begin{subequations}
\label{termAct}
\begin{align}
\displaystyle
S_{cl}[X_1,t_1|X_2,t_2]=&-k_-\,\Bigl[\frac{t_1}{2 \br{R^2+t_1^2}}- \frac{R\,\sqrt{1+\alpha}}{2 \br{R^2+t_1^2}} \,\mathrm{cot}2\Delta_{12}\,\Bigr]X^2_1\label{termX_1X_1}\\
&+k_-\,\Bigl[\frac{t_2}{2 \br{R^2+t_2^2}}+ \frac{R\,\sqrt{1+\alpha}}{2 \br{R^2+t_2^2}} \,\mathrm{cot}2\Delta_{12}\,\Bigr]X^2_2\label{termX_2X_2}\vspace{3mm}\\
&-\Bigl[\frac{k_-\,\sqrt{1+\alpha}\,R}{\sqrt{R^2+t_1^2}\sqrt{R^2+t_2^2}\,\,\mathrm{sin}2\Delta_{12}\,}\Bigr]X_1X_2\\
&-\Bigl[\frac{\beta\,k_{\Theta}}{\sqrt{1+\alpha}\sqrt{R^2+t_1^2}}\mathrm{tan}\Delta_{12}\Bigr]X_1\label{termX_1}\\
&-\Bigl[\frac{\beta\,k_{\Theta}}{\sqrt{1+\alpha}\sqrt{R^2+t_2^2}}\mathrm{tan}\Delta_{12}\Bigr]X_2\label{termX_2}\\
&-\frac{\beta^2\,k_{\Theta}^2}{k_-\br{1+\alpha}^{3/2}\,R}\Bigl(\mathrm{tan}\Delta_{12}\,-\Delta_{12}\Bigr)\label{termOne}\\
&- \frac{m^2}{2 k_-}(t_2-t_1)-\frac{i}{2}\ln\frac{\sqrt{R^2+{t_2}^2}}{\sqrt{R^2+{t_1}^2}}-\frac{k_{\Theta}^2 \Delta_{12}}{k_-\,R\,\sqrt{1+\alpha}}.\label{termDet}
\end{align}
\end{subequations}

Next we consider the ``quantum-mechanical'' prefactor (\ref{sol_f}). With the expression for $S_{cl}$ given by (\ref{Scl_general}), equation (\ref{WKB_pref}) becomes:
\begin{equation}\label{pref_general}
\frac{\partial {\cal A}(t_1,t)}{\partial t}=-\frac12\frac{f_1\dot{h}-h_1\dot{f}}{f_1h-h_1f}\,{\cal A}(t_1,t).
\end{equation}
This leads to the solution
\begin{equation}
{\cal A}(t_1,t_2)=\mathcal{N}\left(R^2+t_1^2\right)^{-1/4} \left(R^2+t_2^2\right)^{-1/4}\,\left|\,\mathrm{sin}2\Delta_{12}\right|^{-1/2}\phi_M,\label{sol_f}
\end{equation}
which contains the Maslov phase $\phi_M$ and a constant normalization factor $\mathcal{N}$. The Maslov phase is piecewise constant away from the focal points (the positions of focal point for $t_2$ are functions of $t_1$). Its value is worked out in appendix \ref{app_maslov}.

We can now fix the normalization of ${\cal A}(t_1,t_2)$ by imposing (\ref{CCnorm}):
\be
{\cal N}=\sqrt{\frac{R \sqrt{1+\alpha}}{2\pi \sqrt{R^2+t_1^2}}}\label{Nresult}
\ee

%%%%%%%%%%%%%%%%%%%%%
\subsubsection{The $R\to 0$ limit}

Now that we have constructed the ``propagator'' $\phi(X_1,t_1|X_2,t_2)$, which provides a basis of mode functions labelled by $X_1$, we can investigate its $R\to 0$ limit, which will only exist for special values of $\alpha$ and $\beta$. 

The first non-trivial condition for the existence of an $R\to0$ limit comes from the prefactor ${\cal A}(t_1,t_2)$. This will vanish identically for $t_1<0$, $t_2>0$ as $R$ is sent to 0 (which would make the field operator vanish identically and manifestly destroy unitarity), unless
\be 
\alpha=N^2-1,
\ee
with an integer $N$, on account of the structure $\sqrt{R/\mathrm{sin}2\Delta_{12}}$ in (\ref{sol_f}) combined with (\ref{Nresult}).

This behavior of the prefactor ${\cal A}(t_1,t_2)$ 
can be naturally understood by inspecting the classical homogeneous solutions (\ref{homSol}). For generic values of $\alpha$ and $\beta$,
in the $R\to 0$ limit, those behave as $f(t)\rightarrow t$ and $h(t)\rightarrow |t|$. Since these two functions are not linearly independent on the negative real axis, one will not be able to specify
arbitrary initial conditions $(X_1$,$V_1)$ for the classical solution at $t_1<0$. Should one try to do so, in the $R\to 0$ limit, the classical trajectory will be kicked away to infinity for all $t>0$.
Correspondingly, all quantum wavepackets will be kicked away to infinity, and the wave function will vanish at all finite values of $X$ for $t>0$, as manifested by the behavior of the prefactor ${\cal A}(t_1,t_2)$. This problem is avoided, however, for special values of $\alpha$: if $\alpha=(2N)^2-1$ (with an integer $N$), the $R\to0$ limit of the two solutions is $f(t)\to \mathrm{sign}(t)$ and $h(t)\rightarrow |t|$; if $\alpha=(2N+1)^2-1$, it is $f(t)\rightarrow t$ and $h(t)\to 1$.

A further condition on $\alpha$ and $\beta$ arises from considering
the $R\to0$ limit of the classical action $S_{cl}[X_1,t_1|X_2,t_2]$.
The problematic terms in the action (\ref{termAct}) are those with coefficients containing $k_{\Theta}^2$, namely:
\begin{equation}
-\frac{\beta^2\,k_{\Theta}^2}{k_-(1+\alpha)^{3/2}\,R}\Bigl(\mathrm{tan}\Delta_{12}-\Delta_{12}\Bigr)-\frac{k_{\Theta}^2 \Delta_{12}}{k_-\,R\,\sqrt{1+\alpha}}.\label{limiting}
\end{equation}
In order to cancel the divergences, we find the following equation for $\beta$:
\begin{equation}
\label{limiting2}
\beta^2=\frac{1+\alpha}{1-\dfrac{\mathrm{tan}\left(\pi\sqrt{1+\alpha}/2\right)}{\pi\sqrt{1+\alpha}/2}}.
\end{equation}
If $\alpha=(2N+1)^2-1$, this condition would make us na\"\i vely
conclude that $\beta=0$. However, a direct inspection of (\ref{limiting}) reveals that it wouldn't make the divergence cancel.
The only other option remaining is
\be
\alpha=(2N)^2-1,\qquad \beta=2N
\ee
($\beta=-2N$ corresponds to the same space written in different coordinates). These are the conditions for existence of an $R\to 0$ limit for a free scalar field dynamics on the generalized null-brane. As one would expect, the values $\alpha=3$ and $\beta=2$ corresponding to the original null-brane do meet these conditions (for $N=1$).

%%%%%%%%%%%%%%%%%%%%%%%
\subsubsection{Mode functions in momentum basis}
\label{modefunctionssection}

To facilitate comparison with earlier work, we shall now derive
momentum basis mode functions using
the position basis mode functions $\phi\br{X_1,t_1|X_2,t_2}$ of section~\ref{2-1/2cl}. The existence of an $R\rightarrow0$ limit will not be affected by such conversion. To obtain the momentum basis mode functions, we take a Fourier transform (with respect to $X_1$) of the ``propagator'' $\phi\br{X_1,t_1|X_2,t_2}$ to convert it to an incoming plane wave basis, cancel the ``free evolution'' by multiplying with $\mathrm{exp}\bbr{i\br{m^2+p^2} t_1 /(2 k_-)}$ (we are simply using the freedom we have in defining the basis of mode functions), and take the $t_1\rightarrow-\infty$ limit (which refers our momentum labels to incoming waves in the infinite past). We shall also omit time-independent overall phase factors. Finally, we shall take the limit $t_1\rightarrow-\infty$:
\begin{equation}
V_{k_-,k_{\Theta},p,m}=\lim\limits_{t_1\rightarrow-\infty}\int_{-\infty}^{\infty}\,\mathrm{d}X_1\,{\cal A}(t_1,t_2)\,\mathrm{exp}\br{-i S_{cl}}\,\mathrm{exp}\br{i p X_1}\mathrm{exp}\br{\frac{i \br{m^2+p^2} t_1}{2 k_-} }\label{modef}
\end{equation}

To be consistent with our notation in the beginning of this section, we shall now switch back to writing $X^+$ instead of $t$ for all space-time quantities. We finally obtain the mode functions
\be
\begin{array}{l}
\dsty V_{k_-,k_{\Theta},p,m} =\sqrt{\frac{2 R N}{|k_-|\left(R^2+(X^+)^2\right)|\,\mathrm{sin}2\Delta\,|}}\,\phi_M\vspace{3mm}\\
\dsty\hspace{1cm}\times\mathrm{exp}\Bigl[\frac{-ik_-\,X^2}{2 \br{R^2+(X^+)^2}}\br{2 N R\,\mathrm{cot}2\Delta\,+X^+}+\frac{i\,X}{\sqrt{R^2+\br{X^+}^2}}\br{k_{\Theta}\mathrm{tan}\Delta+\frac{2 R N p}{\,\mathrm{sin}2\Delta\,}}\vspace{3mm}\\
\dsty\hspace{15mm}+ \frac{i p k_{\Theta}}{k_-}\mathrm{tan}\Delta-\frac{i p^2 R N}{k_-}\,\mathrm{cot}2\Delta\,+ \frac{i k_{\Theta}^2 }{2 k_- N R}\mathrm{tan}\Delta+\frac{i m^2 X^+}{2 k_-}+i k_- X^-+i k_{\Theta}, \Theta\Bigr]\label{modefu}
\end{array}
\ee
where $\Delta=N \mathrm{arctan}(X^+/R) + \pi N/2$.

For $R\rightarrow0$ we obtain:
\begin{align}
\displaystyle
V_{k_-,k_{\Theta},p,m} =&\frac{1}{\sqrt{|k_-X^+|}} \mathrm{exp}\br{\frac{i\pi}{2}\mathrm{sign}(k_-)(2N-1)\theta(X^+)}\nonumber \vspace{2mm}\\
&\times \mathrm{exp}\Bigl[-i p X\,\mathrm{sign}(X^+)+\frac{i p^2 X^+}{k_-}- \frac{i k_{\Theta}^2 }{2 k_- X^+}+\frac{i m^2 X^+}{2 k_-}+i k_- X^-+i k_{\Theta} \Theta\Bigr],
\label{modefR0}
\end{align}
where $\theta(t)$ denotes the Heaviside step function.

In order to compare our mode functions (\ref{modefu}) with those derived in \cite{Liu:2002kb}, we write out the mode functions for $N=1$ explicitly:
\be
\begin{array}{l}
\dsty
V_{k_-,k_{\Theta},p,m}^{(N=1)} =\frac{1}{\sqrt{|k_-X^+|}}\phi_M\mathrm{exp}\Bigl[\frac{-ik_-\,R^2 X^2}{2 X^+ \br{R^2+(X^+)^2}}-\frac{i X}{\sqrt{R^2+(X^+)^2}}\br{k_{\Theta}\frac{R}{X^+}+p \frac{R^2+(X^+)^2}{X^+}}\vspace{3mm}\\
\dsty\hspace{1cm}- \frac{i p k_{\Theta} R}{k_- X^+}-\frac{i p^2}{k_-}\br{\frac{R^2-(X^+)^2}{2 X^+}}- \frac{i k_{\Theta}^2 }{2 k_- X^+}+\frac{i m^2 X^+}{2 k_-}+i k_- X^-+i k_{\Theta} \Theta\Bigr].
\end{array}
\label{modefN1}
\ee
The general expression for the Maslov phase $\phi_M$ is given in appendix~\ref{app_maslov}. For $N=1$, there is only one focal point at $X^+=0$, and the Maslov phase becomes:
\be
\phi_M=\mathrm{exp}\br{\frac{i\pi}{2}\mathrm{sign}(k_-)\,\theta(X^+)}.
\ee

Written in our interpolating coordinates from (\ref{LMScotrafo}), the mode functions of \cite{Liu:2002kb} can be re-expressed as
\be
\begin{array}{l}
\dsty\frac{1}{\sqrt{iX^+}}\mathrm{exp}\Bigl[\frac{i p^+ R^2 X^2}{2 X^+ \br{R^2+(X^+)^2}}-\frac{i\,X}{\sqrt{R^2+(X^+)^2}} \br{J \frac{R^2+(X^+)^2}{R X^+}- n \frac{X^+} {R}}\vspace{3mm}\\
\dsty\hspace{2.5cm}- \frac{iX^+}{2 p^+}\br{\frac{n-J}{R}}^2+\frac{i J^2}{2 p^+ X^+}-\frac{i m^2 X^+}{2 p^+}-i p^+ X^-+i n \Theta\Bigr],
\end{array}
\label{nullbr}
\ee
with $J$, $n$ being the labels used in \cite{Liu:2002kb}.

The equality of the two expressions (up to normalization conventions) is established by recognizing the following identifications:
\be\label{ident}
\begin{array}{l}
\dsty k_-=-p^+\vspace{2mm},\\
\dsty p=\frac{J-n}{R},\vspace{2mm}\\
\dsty k_{\Theta}=n.
\end{array}
\ee
Our results thus agree with those of \cite{Liu:2002kb} for non-zero values of $R$ in the particular case of the null-brane ($\alpha=3$, $\beta=2$). Now we are ready to examine the $R\to 0$ limit (where our choice of coordinates will reveal a peculiar reflection property at the singularity) for the original as well as the generalized null-brane. 

%%%%%%%%%%%%%%%%%%%%%%%%%%%%%%%%%%%%%%%%%%%%%%%%%%%%%%%%%%%%%%%%%%%%%%%%%%%%%%%

\subsection{Qualitative properties of the singular limit}

To recapitulate, we have examined the dynamics of a free scalar field on the following
3-parameter ($\alpha$, $\beta$, $R$) family of backgrounds (``generalized null-brane''):
\be
\begin{array}{l}
\displaystyle ds^2=-2 dX^+ dX^-+ \frac{X^2 R^2 ( \beta^2 - \alpha)}{(R^2+(X^+)^2)^2}(dX^+)^2\,+\frac{2\beta X R}{\sqrt{R^2+(X^+)^2}}dX^+ d\Theta\vspace{3mm}\\
\dsty\hspace{8cm}+\left(R^2+(X^+)^2\right) d\Theta^2+dX^2
\end{array}
\ee
As $R$ goes to 0, all of these geometries (irrespectively of the values of $\alpha$
and $\beta$) reduce (away from $X^+=0$) to the parabolic orbifold times a line:
\be
ds^2=-2 dX^+ dX^-+(X^+)^2 d\Theta^2+dX^2
\label{po}
\ee
What we have found is that the $R\to 0$ limit of the scalar field mode functions exists
only when $\alpha=(2N)^2-1$ and $\beta=2N$, with $N$ being an integer.

In terms of the limiting expression for the mode functions (\ref{modefR0}), we find few
surprises. The result is essentially independent of the values of $\alpha$ and $\beta$
(for those values for which the limit exists) and bears a close resemblance to the mode functions obtained in \cite{Liu:2002kb}. However, the minor discrepancy between our
results and those of \cite{Liu:2002kb} deserves some clarification.

For any $\alpha$ and $\beta$, the metric of the generalized null-brane converges to the metric
of the parabolic orbifold times a line (\ref{po}), which is formally the same as the metric
of the parabolic orbifold written in the $y$-coordinates of \cite{Liu:2002kb}.
Given only the $R=0$ expressions, it may therefore be tempting to identify $(y^+,y^-,y,u)\leftrightarrow (X^+,X^-,\Theta,X)$. With this identification, however, the mode functions are not exactly the same, even for the case of the original null-brane ($\alpha=3$, $\beta=2$). 
The difference between the two is the factor of ${\rm sign}(X^+)$ in front of
the $ipX$ term in the exponential of (\ref{modefR0}). It is important to realize that the difference
between the two sets of mode function does not represent any dynamical distinction.
Rather, it is explained by the difference in the choice of coordinates.

To construct the parabolic orbifold (\ref{po}) as an $R\to 0$ limit of the null-brane,
the authors of \cite{Liu:2002kb} employ their singular $y$-coordinates (this coordinate
system fails at $X^+=0$ even for smooth spaces at non-zero $R$). As a result, they
obtain mode functions without\footnote{Incidentally, the same mode functions are obtained by applying the non-geometrical ``minimal subtraction'' prescription of \cite{qsing} directly to a free scalar field on the parabolic orbifold, without any recourse to the null-brane or its generalizations. See appendix \ref{minsubtr}.} the aforementioned factor ${\rm sign}(X^+)$.

On the other hand, we construct the parabolic orbifold metric (\ref{po}) and the corresponding coordinates as an $R\to 0$ limit of smooth coordinate systems
parametrizing smooth geometries. In this case, the factor of ${\rm sign}(X^+)$
is present and its effect is that the position and velocity in the $X$-direction
for all particles are reflected as they pass through $X^+=0$.

Even though the two sets of mode functions are essentially equivalent (and only
differ by a coordinate choice), one may think of our present parametization as being more accurate. Indeed, it is very natural to demand that, since the singular space is
constructed as an $R\to 0$ limit of smooth resolved geometries, the coordinates
on the singular space should be constructed as an $R\to 0$ limit of smooth coordinate
systems on the smooth resolved geometries (even though, with our present theoretical understanding of space-time singularities, it is not possible to give a systematic
justification to this treatment of coordinate systems). Note that the flip of the $X$-direction for positive $X^+$ cannot be undone by a smooth coordinate transformation,
so it will always be present if the parabolic orbifold metric (\ref{po}) is constructed
as a limit of a smooth coordinate frame on the (generalized) null-brane.

One finds considerably more surprises if one contemplates the properties of those geometrical resolutions for which the singular limit of the scalar field dynamics
exists (rather than merely examining the limiting expressions for the mode functions).

Firstly, looking at the null-brane example of \cite{Liu:2002kb}, one could get the
impression that the singular limit exists because the curvature is identically
zero for any finite $R$ (in a way, one can say that the singularity is never really ``there''). However, this is not the case. The non-vanishing components of the Riemann tensor for our generalized null-brane\footnote{These will obviously vanish for the values of $\alpha$ and $\beta$ corresponding to the original null-brane space-time: $\alpha=3$, $\beta=2$.} geometries are
\be
R_{+X+X}=\frac{R^2(4\alpha-3\beta^2)}{4(R^2+(X^+)^2)^2},\qquad
R_{+\Theta+\Theta}=\frac{R^2(\beta^2-4)}{4(R^2+(X^+)^2)}.
\label{riemann}
\ee
So, generically, the curvature {\it will} blow up around $X^+=0$ as $R$ is sent to 0,
even for those values of $\alpha$ and $\beta$, for which the singular limit of the scalar field dynamics exists ($\alpha=(2N)^2-1$, $\beta=2N$). Note that the Ricci scalar vanishes, so our results
do not depend on the choice of the Ricci scalar coupling of the
scalar field (minimal, conformal, etc., \cite{birrelldavies}).

Furthermore, one could examine the Weyl tensor:
\be
C_{+X+X}=\frac{R^2(\alpha+1-\beta^2)}{2(R^2+(X^+)^2)^2},\qquad
C_{+\Theta+\Theta}=-\frac{R^2(\alpha+1-\beta^2)}{2(R^2+(X^+)^2)}
\label{weyl}
\ee
and notice that it actually {\it does} vanish for all those cases when the limit exists (within the particular family of geometries we have been considering). However, there
are many values of $\alpha$ and $\beta$ for which the Weyl tensor (\ref{weyl}) vanishes,
yet no $R\to 0$ limit of the scalar field dynamics exists. For that reason, conformal
flatness is not likely to constitute an important part in possible explanations for
the existence of the singular limit.

Perhaps the most puzzling feature of our results is the very fact that the limit
appears to exist for a discrete subset of the possible parameter values within
our family of geometries. One could think of this as being an artifact of choosing
our particular slice in the space of all possible geometries (this, however,
would obviously require a fairly delicate coincidence). If, on the other hand,
the feature is generic, it would point to an interesting sort of discreteness
inherent to the dynamics in the near-singular region. This question would certainly
deserve further investigation, even though that would require mathematical machinery
going beyond what has been employed in our present considerations.

%%%%%%%%%%%%%%%%%%%%%%%%%%%%%%%%%%%%%%%%%%%%%%%%%%%%%%%%%%%%%%%%%%%%%%%%%%%%%%%%%%%%%%%%%%%%%%%%%%%%%%%%%%%%%%%%%%%%%%%%%%%%%%%%%%%%%%%%%%%%%%%%%%%%%%%%%%%%%%%%%%%%%%%%%%%%%%%%%%%%%%%%%%%%%%%%%%

\section{The light-like reflector plane}

Our generalized null-brane considerations of the previous section
suggest a very natural simplification. Namely, instead of
(\ref{lineElementGNB}) we shall consider the following metric: 
\be
ds^2=-\frac{\alpha R^2 X^2 }{\br{R^2+(X^+)^2}^2}\br{dX^+}^2\,-2 dX^+
dX^-+dX^2. 
\label{reflectorplane} 
\ee 
This space-time can be classified as a pp-wave geometry (see, for example, \cite{Blau:2003rt}), and, in particular, it can be extended to a 10-dimensional background satisfying Einstein's equation by inclusion of the appropriate
antisymmetric tensor field and dilaton. What makes this particular pp-wave interesting is
that it develops a strong singularity when $R$ is sent to 0 (for example,
the singularity is much more dangerous than the ``weak pp-wave singularities'' of \cite{David:2003vn}). Furthermore, the $R\to0$ limit of the wave equation
in this geometry can be explicitly analyzed.

We shall not present a detailed derivation of the mode functions for (\ref{reflectorplane}), but simply notice that the wave equation in
this background is formally analogous to that on the generalized null-brane (\ref{waeq}), with $\beta=0$, $k_\Theta=0$ and the first term on the right hand side omitted (this term comes from the determinant of the generalized null-brane metric). The expression for the position basis mode functions $\phi(X_1,t_1|X_2,t_2)={\cal A}(t_1,t_2)\exp[-iS_{cl}(X_1,t_1|X_2,t_2)]$ then follows from (\ref{termAct}, \ref{sol_f}, \ref{Nresult}):
\begin{subequations}
\label{termAct_rpl}
\begin{align}
\displaystyle
S_{cl}[X_1,t_1|X_2,t_2]=&-k_-\,\Bigl[\frac{t_1}{2 \br{R^2+t_1^2}}- \frac{R\,\sqrt{1+\alpha}}{2 \br{R^2+t_1^2}} \,\mathrm{cot}2\Delta_{12}\,\Bigr]X^2_1\label{termX_1X_1_rpl}\\
&+k_-\,\Bigl[\frac{t_2}{2 \br{R^2+t_2^2}}+ \frac{R\,\sqrt{1+\alpha}}{2 \br{R^2+t_2^2}} \,\mathrm{cot}2\Delta_{12}\,\Bigr]X^2_2\label{termX_2X_2_rpl}\vspace{3mm}\\
&-\Bigl[\frac{k_-\,\sqrt{1+\alpha}\,R}{\sqrt{R^2+t_1^2}\sqrt{R^2+t_2^2}\,\,\mathrm{sin}2\Delta_{12}\,}\Bigr]X_1X_2 - \frac{m^2}{2 k_-}(t_2-t_1)\label{termDet_rpl}
\end{align}
\end{subequations}

\begin{equation}
{\cal A}(t_1,t_2)=\biggl(\frac{2 \pi}{R \sqrt{1+\alpha}}\sqrt{R^2+t_1^2}\sqrt{R^2+t_2^2}\,|\,\mathrm{sin}2\Delta_{12}\,|\biggr)^{-\frac{1}{2}}\phi_M\label{sol_f_rpl}
\end{equation}
(with $\Delta_{12}=\sqrt{1+\alpha}\left(\mathrm{arctan}(t_2/R)-\mathrm{arctan}(t_1/R)\right)/2$ and $\phi_M$ being the appropriate Maslov phase).

The $R\to0$ limit of $\phi(X_1,t_1|X_2,t_2)$ exists if $\alpha=K^2-1$ (with $K$ being an integer) and equals 
\be
\begin{array}{l}
\dsty\phi(X_1,t_1|X_2,t_2)=\frac{1}{\sqrt{2 \pi|t_2-t_1|}}\,\phi_M\vspace{3mm}\\
\dsty\hspace{2cm}\times\exp\left[\frac{ik_-}2\left(X_2-\left(\mathrm{sign}(t_1)\mathrm{sign}(t_2)\right)^{K+1}X_1\right)^2+ \frac{im^2}{2 k_-}(t_2-t_1)\right].
\end{array}
\label{rfp_phi}
\ee

If $K$ is odd, the above expression merely represents free motion
on Minkowski space. To verify this statement, one can simply check that $\phi(X_1,t_1|X_2,t_2)$ solves the Minkowski space wave equation written in light cone coordinates:
\be
-i\dot{\phi}=-\frac{\partial^2_X\phi}{2 k_-}+\frac{m^2}{2 k_-}\phi
\ee
Despite the strength of the singularity in the $R\to 0$ limit, the free scalar field dynamics actually becomes identical to that on a completely flat space.

If $K$ is even, the motion is still free for all positive and all negative $t$ (read $X^+$). However, as the particles pass through
$X^+=0$, their positions and velocities in the $X$-direction are reflected (note the $\left(\mathrm{sign}(t_1)\mathrm{sign}(t_2)\right)^{K+1}$ structure in (\ref{rfp_phi})). This reflection is similar to the one happening for the case of the generalized null-brane, but it occurs on a simpler
space-time geometry. The mode functions corresponding to (\ref{rfp_phi}) can be easily derived by means of a Fourier transformation, analogously to section \ref{modefunctionssection}.

Because of the property we have just described, we call the space-time (\ref{reflectorplane}) with $\alpha=(2N)^2-1$, where $N$ is an integer, the {\it light-like reflector plane}. It is an extremely simple family of pp-wave geometries developing a strong singularity at $X^+=0$ when $R$ is sent to 0. Furthermore, at least for free propagation in this background, the singular limit is manifestly well-defined, and includes a curious light-like object reflecting the positions and velocities of all particles passing through it.
Given its simplicity, the background presented in this section may be worth studying as a toy model for light-like singularities in both perturbative string theory and matrix theory contexts.

%%%%%%%%%%%%%%%%%%%%%%%%%%%%%%%%%%%%%%%%%%%%%%%%%%%%%%%%%%%%%%%%%%%%%%%%%%%%%%%%%%%%%%%%%%%%%%%%%%%%%%%%%%%%%%%%%%%%%%%%%%%%%%%%%%%%%%%%%%%%%%%%%%%%%%%%%%%%%%%%%%%%%%%%%%%%%%%%%%%%%%%%%%%%%%%%%%

\section{Conclusions}

In this paper, we have noted the relevance of Hamiltonians 
with singular time dependences involving multiple operator structures
for the problem of geometrical resolution of singular space-times.
We have given a general review of the quantum dynamics corresponding
to this type of Hamiltonians, and emphasized some important simplifications,
which can occur if the Hamiltonian possesses a finite dimensional
{\it dynamical group}.

Turning further to a particular case of interest, we considered
a 2-parameter generalization of the null-brane space-time, and addressed
the question of the singular limit of a free scalar field dynamics
on this background. Surprisingly, this limit happened to exist for a
{\it discrete} subset of the possible values of the two parameters.
The limiting mode functions are closely related to those previously obtained for the null-brane \cite{Liu:2002kb}. We have opted
for an accurate coordinatization of the singular limit of our spaces,
based on taking a limit of smooth coordinate systems on the smooth
geometrical resolutions. In contrast to those employed in \cite{Liu:2002kb}, our coordinate system reveals a peculiar
``reflection'' property of the generalized (as well as the original)
null-brane space-times.

Our analysis of the generalized null-brane suggested a natural
simplification, and we termed this simplified geometry the {\it light-like reflector plane}. This space-time is a fairly simple pp-wave,
whose singular limit is a light-like plane reflecting the positions and velocities of all particles as they pass through it. Due to the simplicity of its singular structure, the light-like reflector plane may turn out to be an interesting toy model for studying light-like space-time singularities in perturative string theory and matrix theory.

%%%%%%%%%%%%%%%%%%%%%%%%%%%%%%%%%%%%%%%%%%%%%%%%%%%%%%%%%%%%%%%%%%%%%%%%%%%%%%%%%%%%%%%%%%%%%%%%%%%%%%%%%%%%%%%%%%%%%%%%%%%%%%%%%%%%%%%%%%%%%%%%%%%%%%%%%%%%%%%%%%%%%%%%%%%%%%%%%%%%%%%%%%%%%%%%%%%

\section*{Acknowledgments}
We would like to thank J.~Maldacena for posing a question which has served as a point of departure for this work, and M.~Gaberdiel and F.~Larsen for discussions. The research presented here has been supported in part by the Belgian Federal Science Policy Office through the Interuniversity Attraction Pole IAP VI/11, by the European Commission FP6 RTN programme MRTN-CT-2004-005104 and by FWO-Vlaanderen through project G.0428.06.

%%%%%%%%%%%%%%%%%%%%%%%%%%%%%%%%%%%%%%%%%%%%%%%%%%%%%%%%%%%%%%%%%%%%%%%%%%
\appendix

\section{Minimal subtraction for the parabolic orbifold}
\label{minsubtr}

It is instructive to compare our derivations of section 3 (based on geometrical regularizations of the singularity) with what we would have obtained by applying the ``minimal subtraction'' prescription of \cite{qsing} directly to the parabolic orbifold.

We write the action for a massive scalar field in the parabolic orbifold using the metric (\ref{metricRosenPO}):
\begin{equation}
S =\int \mathrm{d}y\,\mathrm{d}y^+\,\mathrm{d}y^-\,\mathrm{d}z\,|y^+| \br{\partial_+ \phi \partial_- \phi-\,\frac{(\partial_y\phi)^2}{2\,(y^+)^2}-\,\frac{m^2\phi^2}{2}}.
\end{equation}
We decompose $\phi$ into Fourier modes along $y^-$ and $y$ (with the condition $\phi_{l,k_y}^*=\phi_{-l,-k_y}$):
\begin{equation}
\phi=\frac{1}{2\pi}\sum_{k_y}\int dl \phi_{l,k_y} \mathrm{exp}(ily^-+ik_yy).
\end{equation}
Now the action can be rewritten as
\begin{equation}
S =\sum_{k_y}\int \mathrm{d}y^+\,dl\,|y^+| \Bigl[il \br{\phi_{l,k_y}\partial_+ \phi_{l,k_y}^* -\phi_{l,k_y}^*\partial_+\phi_{l,k_y}}-\br{\frac{k_y^2}{(y^+)^2}+\,m^2}\phi_{l,k_y}\phi_{l,k_y}^*\Bigr]
\end{equation}
The equations of motion are
\begin{equation}
2il\partial_+\phi_{l,k_y}+\frac{il\phi_{l,k_y}}{y^+}+\br{\frac{k_y^2}{(y^+)^2}+\,m^2}\phi_{l,k_y}=0.
\end{equation}
One can deal with the constraints due to the first order nature of the light-cone formalism by choosing $\tilde{\phi}_{l,k_y}=\sqrt{il y^+}\phi_{l,k_y}$ as the canonical coordinate and $\tilde{\pi}_{l,k_y}=\sqrt{il y^+} \phi_{l,k_y}^*$ as its conjugate momentum. We obtain the Hamiltonian
\begin{equation}
H=\sum_{k_y} \int \mathrm{d}y^+\,dl\, \frac{1}{2il} \br{\frac{k_y^2}{(y^+)^2}+\,m^2} \tilde{\pi}_{l,k_y}\tilde{\phi}_{l,k_y}.\label{ham}
\end{equation}
We now apply the ``minimal subtraction'' scheme of \cite{qsing} to the singular time ($y^+$) dependence in (\ref{ham}):
\begin{equation}
\frac{1}{(y^+)^2}\rightarrow \frac{(y^+)^2-\epsilon^2}{\br{(y^+)^2+\epsilon^2}^2}.
\end{equation}
(One could in principle add a (resolved) $\de$-function with an arbitrary coefficient on the right hand side, but we shall not make use of this freedom for the sake of brevity.) The solution for $\tilde{\phi}$ reads:
\begin{equation}
\tilde{\phi}_{l,k_y} \propto \mathrm{exp}\br{-\frac{m^2}{2il} y^+ + \frac{k_y^2}{2il}\frac{y^+}{(y^+)^2+\epsilon^2}}.
\end{equation}
We can return to the original $\phi_{l,k_y}$ and write the scalar field mode functions as:
\begin{equation}
\psi_{l,k_y,m^2}(y^+,y^-,y)\propto \frac{1}{\sqrt{2il y^+}} \mathrm{exp}\br{-\frac{m^2}{2il} y^+ + \frac{k_y^2}{2il}\frac{y^+}{(y^+)^2+\epsilon^2}\,+i k_y y\,+\,i l y^-}.
\label{mofu}
\end{equation}

In order to compare with the mode functions of \cite{Liu:2002kb}, we identify $k_y=J$ and $p^+=-p_-=-l$. Furthermore, the mode functions
of \cite{Liu:2002kb} are derived for the parabolic orbifold times a line, whereas the mode functions (\ref{mofu}) refer to the parabolic
orbifold proper. To compensate for this difference, one should set
the momentum along the extra line in \cite{Liu:2002kb} to zero
(which amounts to imposing $n=J$ in the notation of that paper).
Thereafter, the two sets of mode functions agree.

Note that the agreement is largely accidental. The mode function
of \cite{Liu:2002kb} are derived via a geometrical regularization,
but they are written in the singular $y$-coordinates (see further discussion in section 3.3). The minimal subtraction mode function
are obtained through a regularization procedure that does not
admit a geometrical interpretation.

%%%%%%%%%%%%%%%%%%%%%%%%%%%%%%%%%%%%%%%%%%%%%%%%%%%%%%%%%%%%%%%%%%%%%%%%%%%%%%%%%%%%%%%%%%%%%%%%%%%%%%%%%%%%%%%%%%%%%%%%%%%%%%%%%%%%%%%%%%%%%%%%%%%%%%%%%%%%%%%%%%%%%%%%%%%%%%%%%%%%%%%%%%%%

\section{Focusing properties of the wave equation and the Maslov phase\label{app_maslov}}

In evolution of classical dynamical systems, it often happens
that {\it all} classical trajectories starting at $(X_1,t_1)$
reach the same point $X^*(X_1)$ at the moment $t^*(t_1)$
(irrespectively of their initial velocity $V_1$). Under such circumstances, one speaks of $t^*(t_1)$ as being a classical
{\it focal point} of the evolution.

If $t^*(t_1)$ is such a focal point, the classical action
$S_{cl}[X_1,t_1|X_2,t^*(t_1)]$ will diverge unless $X_2=X^*(X_1,t_1)$
(for, did it not, there would have been classical trajectories
connecting $(X_1,t_1)$ and $(X_2,t^*(t_1))$ for $X_2\ne X^*(X_1,t_1)$,
in contradiction with the definition of a focal point). As we already
remarked in section \ref{2-1/2cl}, if one pursues a semiclassical construction of the quantum-mechanical mode functions, such singular behavior of the classical action introduces formal complications in equation (\ref{WKB_pref}). A general recipe for handling this type of singular features can be given \cite{rohmer,maslov}. However, in our present context, it will be more practical to analyze the relevant
solution for our particular form of the classical action. 

To this end, we shall rewrite (\ref{Scl_general}) in the following form:
\begin{align}
S_{cl}[X_1,t_1|X,t]&=\frac{k_-}{2}\bbr{\frac{f_1\dot{h}-h_1\dot{f}}{f_1h-h_1f}} \br{X - X^*(X_1,t_1,t)}^2\;+\;\cdots\\
&=\frac{k_-}{2}\frac{\partial}{\partial t}\ln\bbr{f_1 h-h_1 f} \br{X - X^*(X_1,t_1,t)}^2\;+\;\cdots,
\end{align}
where the dots represent contributions non-singular at $t=t^*(t_1)$. A focal point $X^*(X_1,t_1,t^*(t_1))$ is reached whenever
\be
f_1h(t^*)-h_1f(t^*)\equiv f(t_1)h(t^*)-h(t_1)f(t^*)=0.
\ee
At the same time, equation (\ref{pref_general}) for the prefactor
${\cal A}(t_1,t)$ can be solved on the left and on the right of
the focal point $t^*(t_1)$ (even though constructing a solution
at $t=t^*(t_1)$ na\"\i vely would be problematic on account
of the singularity on the right hand side of (\ref{pref_general})):
\be
\begin{array}{l}
\dsty{\cal A}(t_1,t)={\cal N}_<\left|f_1h(t)-h_1f(t)\right|^{-1/2}\qquad(t<t^*(t_1)),\vspace{2mm}\\
\dsty{\cal A}(t_1,t)={\cal N}_>\left|f_1h(t)-h_1f(t)\right|^{-1/2}\qquad(t>t^*(t_1))
\end{array}
\ee
(with ${\cal N}_<$ and ${\cal N}_>$ being complex constants).

Armed with these relations, we can examine the behavior of the
entire wave function $\phi(X_1,t_1|X,t)={\cal A}(t_1,t) \mathrm{exp}\br{-i S_{cl}\bbr{X_1,t_1|X,t}}$ in the vicinity of a focal point $t^*(t_1)$. Generically assuming that $f_1h(t)-h_1f(t)$
has a simple zero at the focal point, $f_1h(t)-h_1f(t)\propto t-t^*(t_1)$, and keeping in mind that
\begin{equation}
\lim\limits_{\lambda\to\infty} \sqrt{\frac{|\lambda|}{\pi}}\,\mathrm{exp}\br{ - \frac{i \pi}{4}\mathrm{sign}(\lambda)}\mathrm{exp}\br{i \lambda x^2}\;=\; \de(x),\label{GIphase}
\end{equation}
we conclude that
\be
\begin{array}{l}
\dsty \lim\limits_{t\to (t^*(t_1))_-}\phi(X_1,t_1|X,t)={\cal A}_<\de(X-X^*(X_1,t_1)),\vspace{2mm}\\
\dsty\lim\limits_{t\to (t^*(t_1))_+}\phi(X_1,t_1|X,t)={\cal A}_>\de(X-X^*(X_1,t_1)),
\end{array}
\label{focallimits}
\ee
with
\be
\frac{{\cal A}_<}{{\cal A}_>}=\frac{{\cal N}_<\,\mathrm{exp}\br{i\pi\,\mathrm{sign}(k_-)/4}}{{\cal N}_>\mathrm{exp}\br{-i\pi\,\mathrm{sign}(k_-)/4}}
\ee
If we further demand that the limits in (\ref{focallimits}) should
be the same (this automatically ensures that any convolution of $\phi(X_1,t_1|X,t)$ with a smooth wave packet is continuous across the focal point), we conclude that
\begin{equation}
\frac{{\cal N}_>}{{\cal N}_<}=\mathrm{exp}\br{\frac{i\pi}{2}\mathrm{sign}(k_-)}.\label{Mas1}
\end{equation}
When there are many focal points $t^*_\ell(t_1)$, each of them will give a contribution,
and the resulting wave function can be written as
\be
\phi(X_1,t_1|X,t)={\cal N}\phi_M \left|f_1h(t)-h_1f(t)\right|^{-1/2}\mathrm{exp}\br{-i S_{cl}\bbr{X_1,t_1|X,t}}
\ee
with a constant normalization factor ${\cal N}$ and the {\it Maslov phase} $\phi_M$ of the form
\begin{equation}
\phi_M=\mathrm{exp}\br{\frac{i\pi}{2}\mathrm{sign}(k_-)\,\sum\limits_\ell\,\theta(t-t^*_\ell)}
\end{equation}
($\theta(t)$ being the Heaviside step function).

We now turn to our specific case for which the relevant term in the classical action (\ref{termAct}) near a focal point is given by:
\begin{equation}
S_{cl}[X_1,t_1|X,t]\simeq k_-\,\Bigl[\frac{t}{2 \br{R^2+t^2}}+ \frac{R\,\sqrt{1+\alpha}}{2 \br{R^2+t^2}} \,\mathrm{cot}2\Delta_{1t}\,\Bigr]X^2+\cdots
\end{equation}
The focal points are the poles of $\mathrm{cot}2\Delta_{1t}$:
\begin{equation}
t^*=\frac{t_1 + R\,\mathrm{tan}(\pi\ell/\sqrt{1+\alpha})}{1- \mathrm{tan}(\pi\ell/\sqrt{1+\alpha})\,t_1/R}\;,\hspace{5mm}\ell\in\mathbb{Z}
\end{equation}
The value of $\alpha$ determines the number of focal points. We will restrict our attention to the case $\alpha=(2N)^2-1$ relevant for our present investigation, and we obtain $\ell\in\{1,\ldots,2N-1\}$, i.e., $2N-1$ focal points.

In the $R\rightarrow0$ limit, all the $2N-1$ focal points are squeezed into $t=0$. This means that there will be one large phase jump from $t<0$ to $t>0$. We thus obtain the following expression for the Maslov phase in the $R\to0$ limit:
\begin{equation}
\phi_M=\mathrm{exp}\br{\frac{i\pi}{2}\mathrm{sign}(k_-)\br{2N-1}\theta(t)}.
\end{equation}

%%%%%%%%%%%%%%%%%%%%%%%%%%%%%%%%%%%%%%%%%%%%%%%%%%%%%%%%%%%%%%%%%%%%%%%%%%%%%%%%%%%%%%%%%%%%%%%%%%%%%%%%%%%%%%%%

%%%%%%%%%%%%%%%%%%%%%%%%%%%%%%%%%%%%%%%%%%%%%%%%%%%%%%%%%%%%%%%%%%%

\begin{thebibliography}{99}

\bibitem{ben1}
  B.~Craps, S.~Sethi and E.~P.~Verlinde,
  ``A matrix big bang,''
  JHEP {\bf 0510} (2005) 005
  {\tt [arXiv:hep-th/0506180]};
   B.~Craps, A.~Rajaraman and S.~Sethi,
  ``Effective dynamics of the matrix big bang,''
  Phys.\ Rev.\  D {\bf 73} (2006) 106005
  {\tt [arXiv:hep-th/0601062]}.

\bibitem{hertog}
  T.~Hertog and G.~T.~Horowitz,
  ``Towards a big crunch dual,''
  JHEP {\bf 0407} (2004) 073
  {\tt [arXiv:hep-th/0406134]};
``Holographic description of AdS cosmologies,''
  JHEP {\bf 0504} (2005) 005
  {\tt [arXiv:hep-th/0503071]};
   N.~Turok, B.~Craps and T.~Hertog,
  ;``From Big Crunch to Big Bang with AdS/CFT,''
  {\tt arXiv:0711.1824 [hep-th]};
B.~Craps, T.~Hertog and N.~Turok,
  ``Quantum Resolution of Cosmological Singularities using AdS/CFT,''
  {\tt arXiv:0712.4180 [hep-th]}.

\bibitem{robbins}
  D.~Robbins and S.~Sethi,
  ``A matrix model for the null-brane,''
  JHEP {\bf 0602} (2006) 052
  [arXiv:hep-th/0509204];
  E.~J.~Martinec, D.~Robbins and S.~Sethi,
  ``Toward the end of time,''
  JHEP {\bf 0608} (2006) 025
  {\tt[arXiv:hep-th/0603104]}.

\bibitem{FigueroaO'Farrill:2001nx}
  J.~Figueroa-O'Farrill and J.~Simon,
  ``Generalized supersymmetric fluxbranes,''
  JHEP {\bf 0112} (2001) 011
  {\tt[arXiv:hep-th/0110170]}.

\bibitem{Horowitz:1991ap}
  G.~T.~Horowitz and A.~R.~Steif,
  ``Singular string solutions with nonsingular initial data,''
  Phys.\ Lett.\  B {\bf 258} (1991) 91.

\bibitem{miaoli}
M.~Li,
  ``A class of cosmological matrix models,''
  Phys.\ Lett.\  B {\bf 626} (2005) 202
  {\tt [arXiv:hep-th/0506260]}.

\bibitem{Liu:2002ft}
  H.~Liu, G.~W.~Moore and N.~Seiberg,
  ``Strings in a time-dependent orbifold,''
  JHEP {\bf 0206} (2002) 045
  {\tt [arXiv:hep-th/0204168]}.


\bibitem{Liu:2002kb}
  H.~Liu, G.~W.~Moore and N.~Seiberg,
  ``Strings in time-dependent orbifolds,''
  JHEP {\bf 0210} (2002) 031
  {\tt [arXiv:hep-th/0206182]}.

\bibitem{Lawrence:2002aj}
  A.~Lawrence,
  ``On the instability of 3D null singularities,''
  JHEP {\bf 0211} (2002) 019
  {\tt[arXiv:hep-th/0205288]}.

\bibitem{Horowitz:2002mw}
  G.~T.~Horowitz and J.~Polchinski,
  ``Instability of spacelike and null orbifold singularities,''
  Phys.\ Rev.\  D {\bf 66} (2002) 103512
  {\tt[arXiv:hep-th/0206228]}.

\bibitem{Berkooz:2002je}
  M.~Berkooz, B.~Craps, D.~Kutasov and G.~Rajesh,
  ``Comments on cosmological singularities in string theory,''
  JHEP {\bf 0303} (2003) 031
  {\tt[arXiv:hep-th/0212215]}.

\bibitem{McGreevy:2005ci}
  J.~McGreevy and E.~Silverstein,
  ``The tachyon at the end of the universe,''
  JHEP {\bf 0508}, 090 (2005)
  {\tt[arXiv:hep-th/0506130]}.

\bibitem{qsing}
  B.~Craps and O.~Evnin,
  ``Quantum evolution across singularities,''
  {\tt arXiv:0706.0824 [hep-th]}.

\bibitem{magnus}
W.~Magnus, ``On the exponential solution of differential equations for a linear operator,'' Commun. Pure Appl. Math. {\bf 7} (1954) 649.

\bibitem{hall}
B.~C.~Hall, {\it Lie Groups, Lie Algebras, and Representations}, Springer-Verlag: New York, 2003.

\bibitem{zhengfenggilmore}W.-M.~Zhang, D.~H.~Feng, and R.~Gilmore, ``Coherent states: theory and some applications,'' Rev. Mod. Phys. {\bf 62} (1990) 867.

\bibitem{malkinmanko}I.~A.~Malkin, V.~I.~Man'ko. {\it Dinamicheskie simmetrii i kogerentnye sostojanija}, Nauka (1979).

\bibitem{yuen}H.~P.~Yuen, ``Two-photon coherent states of the radiation field,'' Phys. Rev. A {\bf 13} 2226.

\bibitem{dodonovmankorosa}V.~V.~Dodonov, V.~I.~Man'ko, and L.~Rosa, ``Quantum singular oscillator as a model of a two-ion trap'', Phys. Rev. A {\bf 57} (1998) 2851 {\tt [arXiv:quant-ph/9712028]}.

\bibitem{rohmer}H.~R\"omer, {\it Theoretical Optics}, Wiley-VCH (2004).

\bibitem{maslov}V.~P.~Maslov, {\it Asimptoticheskie metody i teoriya vozmushchenij}, Nauka (1988).

\bibitem{zaitsev}A.~D.~Polyanin, V.~F.~Zaitsev, {\it Handbook of exact solutions for ordinary differential equations}, Chapman and Hall/CRC (2003).

\bibitem{Fabinger:2002kr}
  M.~Fabinger and J.~McGreevy,
  ``On smooth time-dependent orbifolds and null singularities,''
  JHEP {\bf 0306} (2003) 042
  {\tt [arXiv:hep-th/0206196]}.

\bibitem{RS}
D.~Robbins and S.~Sethi, unpublished.

\bibitem{birrelldavies}
N.~D.~Birrell and P.~C.~W.~Davies,
  {\it Quantum Fields In Curved Space},\\
  Cambridge University Press (1982)

\bibitem{Blau:2003rt}
  M.~Blau, M.~O'Loughlin, G.~Papadopoulos and A.~A.~Tseytlin,
  ``Solvable models of strings in homogeneous plane wave backgrounds,''
  Nucl.\ Phys.\  B {\bf 673} (2003) 57
  {\tt [arXiv:hep-th/0304198]}.
  
\bibitem{David:2003vn}
  J.~R.~David,
  ``Plane waves with weak singularities,''
  JHEP {\bf 0311} (2003) 064
  {\tt [arXiv:hep-th/0303013]}.

\end{thebibliography}
\end{document}